\documentclass[onecolumn,showpacs,preprintnumbers,amsmath,amssymb]{revtex4}

\usepackage{graphicx}
\usepackage{dcolumn}
\usepackage{bm}

\begin{document}

\preprint{APS/123-QED}

\title{A Special Relativistic Liouville Equation Exists}

\author{Jose A. Magpantay}
\email{jose.magpantay11@gmail.com}
\affiliation{Quezon City, Philippines}%
\date{\today}

\begin{abstract}
In a previous paper, the author asked the question 'Does a Special Relativistic Liouville Equation Exist?'. In this paper, I give an affirmative answer. In 8N phase space, a Hamiltonian is derived by breaking the reparametrization symmetry of the single Lorentz invariant mathematical time introduced, which defines the evolution of all phase space variables.  
\end{abstract}

\pacs{Valid PACS appear here}
\maketitle


1. In a previous paper, the author raised the question 'Does a Special Relativistic Liouville Equation Exist' \cite{Magpantay1}. This question is raised because a special relativistic treatment of a system of many particles, typically of the order of $ 10^{23} $, suffers from two problems (a) what time to use since each particle will have its own relativistic time, and (b) there is a problem of existence of a Hamiltonian for interacting many particles \cite{Currie}, \cite{Cannon}, \cite{Leutwyler}. The surprising result of my previous paper is that even for non-interacting relativistic many particles, there is no relativistic Hamiltonian once use is made of a single time for all particles. The reason is that a single time, which must be a Lorentz invariant to maintain Lorentz symmetry, will result in an action that has time reparametrization symmetry. This results in a constraint, which gives a zero Hamiltonian, in the same way as the diffeomorphism symmetry in general relativity gives a constraint resulting in a zero Hamiltonian. 

2. The action for the many interacting relativistic particles is given by
\begin{subequations}\label{1}
\begin{gather}
S_{r} = \sum_{a} \int dt_{a} L_{r}(a), \label{first}\\
L_{r}(a) = (-mc^{2})\left( 1 - \frac{1}{c^{2}} \dfrac{d\vec{x}_{a}}{dt_{a}} \cdot \dfrac{d\vec{x}_{a}}{dt_{a}} \right)^{\frac{1}{2}} - \frac{1}{2} \sum_{b \neq a} \int dt_{b} G_{4}\left(  \vec{x}_{a} - \vec{x}_{b}; t_{a} - t_{b} \right),
\end{gather}
\end{subequations}
where $ t_{a}, a = 1, ... N $ are the relativistic times of each particle. The first term of $ L_{r}(a) $, which is Lorentz invariant, is the free relativistic particles term.  The second is the relativistic two-body interaction term given by the prescription \cite{Magpantay2}
\begin{subequations}\label{2}
\begin{gather}
n^{2}(\vert \vec{k} \vert^{2}) = \left[ \int d^{3}x V(\vert \vec{x} \vert) \exp {i \vec{k} \cdot \vec{x} } \right]^{-1}, \label{first}\\
G_{4}(\vec{x}(t) - \vec{x'}(t'); t - t') = \int d^{3}k dk_{0} \dfrac{\exp {[ik_{0}(t - t') - i\vec{k} \cdot (\vec{x} - \vec{x'})]} }{n^{2}( \vert \vec{k} \vert^{2} - \frac{k_{0}^{2}}{c^{2}} )}, \label{second}\\
V_{r}(\vert \vec{x} - \vec{x'} \vert; t) = \frac{1}{2} \int dt' G_{4}(\vec{x}(t) - \vec{x'}(t'); t - t'). 
\end{gather}
\end{subequations} 
The $ G_{4} $, by construction is Lorentz invariant but the time integrations destroy the Lorentz invariance. However, it gives known results for relativistic potentials.
  
The canonical procedure makes use of equal time Poisson brackets (or ETCs), thus there is a need to introduce a single time, which at this stage can be considered as an artifact time, a mathematical time and not necessarily measured by a clock. Let us postulate that each space-time coordinate is dependent on this universal time t, i.e., 
\begin{equation}\label{3}
x_{a \mu}(t)= (ct_{a}(t),\vec{x}_{a}(t)),
\end{equation}
for $ a = 1,..., N $ and $ \mu = 0, 1, 2, 3 $. Equation (1) then becomes 
\begin{subequations}\label{4}
\begin{gather}
S_{r} = \int dt L_{r}, \label{first}\\
\begin{split}
L_{r}& = \sum_{a} \Big\lbrace (-mc) \left( \eta_{\mu \nu} \dfrac{dx_{a \mu}}{dt}  \dfrac{dx_{a \nu}}{dt} \right)^{\frac{1}{2}} \\
& \quad - \frac{1}{2c^{2}} \dfrac{dx_{a 0}}{dt} \sum_{b \neq a} \int dt' \dfrac{dx_{b 0}}{dt'} G_{4}( \vec{x}_{a} - \vec{x}_{b}; \frac{1}{c} ( x_{a 0}(t) - x_{b 0}(t') ) )\Big\rbrace,
\end{split}
\end{gather}
\end{subequations}
where the subscript r in above means special relativistic. The Lagrangian is now in terms of a single time t, except for the time-delayed two-body interaction term given by a single time delay t'. This time must be a Lorentz scalar so the Lorentz properties of equation (1) are also respected by equation (4). 

3. As shown in \cite{Magpantay1}, the canonical procedure acted on equation (4) yields a zero Hamiltonian in 8N phase space, which led to the claim that there is no special relativistic Liouville equation. The root of this result is the reparametrization symmetry, i.e., under $ t \rightarrow t' = t'(t) $, the action $ S_{r} $ remains the same. But since it is rather hard to accept that a fundamental equation in statistical mechanics will be banned by special relativity, there must be a way out of this problem. 

In the same paper, a way out was suggested in 6N phase space. Start from the action given by equation (2), then use the time of one of the particles, say $ t_{1} $, as the physical time. Express the times of all the other particles $ t_{a} $, for $ a = 2,... N $ making use of the time-delay from $ G_{4}(\vec{x}_{1} - \vec{x}_{a}; t_{1} - t_{a}) $. Since there is now one time $ t_{1} $, the canonical procedure can now be applied on equation (1). But this is not easy to consistently implement.

In this paper, I break instead the reparametrization symmetry to derive a Hamiltonian in 8N phase space.  Since there will be many ways to do this, it seems that the Hamiltonian will not be unique. And indeed, as I will show that by using two ways of breaking the symmetry, I indeed get two Hamiltonians. But which one is correct or at least better? 

4. If equation (4) is replaced by
\begin{subequations}\label{5}
\begin{gather}
S_{r,\alpha} = \int dt L_{r,\alpha}, \label{first}\\
\begin{split}
L_{r,\alpha}& = \sum_{a} \Big\lbrace (-mc) \left( \eta_{\mu \nu} \dfrac{dx_{a \mu}}{dt}  \dfrac{dx_{a \nu}}{dt} \right)^{\frac{1}{\alpha}} \\
& \quad - \frac{1}{2c^{2}} \dfrac{dx_{a 0}}{dt} \sum_{b \neq a} \int dt' \dfrac{dx_{b 0}}{dt'} G_{4}( \vec{x}_{a} - \vec{x}_{b}; \frac{1}{c} ( x_{a 0}(t) - x_{b 0}(t') ) )\Big\rbrace,
\end{split}
\end{gather}
\end{subequations}
where the subscript $ \alpha $ in above refers to the parameter in reparametrization symmetry breaking. Equation (5) differs from equation (4) by replacing $ \frac{1}{2} $ with $ \frac{1}{\alpha} $ in the kinetic term. This will make the action, in particular the kinetic term, not invariant, under reparametrization $ t \rightarrow t' = t'(t) $ as given by
\begin{equation}\label{6}
\int dt' \sum_{a} (-mc) \left( \eta_{\mu \nu} \dfrac{dx_{a \mu}}{dt'}  \dfrac{dx_{a \nu}}{dt'} \right)^{\frac{1}{\alpha}} = \int dt \left[ \dfrac{dt'}{dt} \right]^{1-\frac{2}{\alpha}} \sum_{a} (-mc) \left( \eta_{\mu \nu} \dfrac{dx_{a \mu}}{dt}  \dfrac{dx_{a \nu}}{dt} \right)^{\frac{1}{\alpha}}.
\end{equation}
Note, invariance under reparametrization when $ \alpha = 2 $. 

Doing the canonical procedure without the interaction term (put $ G_{4} = 0 $), the velocities are 
\begin{subequations}\label{7}
\begin{gather}
\dfrac{dx_{a0}}{dt} = \left( \frac{-1}{mc} \right)^{\frac{\alpha}{2-\alpha}} \left( \frac{\alpha}{2}\right)^{\frac{\alpha}{2-\alpha}} \left( \pi_{a}^{2} - \vec{p}_{a} \cdot \vec{p}_{a} \right)^{\frac{\alpha-1}{2-\alpha}} \pi_{a},\label{first}\\
\dfrac{d\vec{x}_{a}}{dt} = - \left( \frac{-1}{mc} \right)^{\frac{\alpha}{2-\alpha}} \left( \frac{\alpha}{2} \right)^{\frac{\alpha}{2-\alpha}} \left( \pi_{a}^{2} - \vec{p}_{a} \cdot \vec{p}_{a} \right)^{\frac{\alpha-1}{2-\alpha}} \vec{p}_{a}.
\end{gather}
\end{subequations}
This equation clearly shows the singular limit of $ \alpha \rightarrow 2 $, the velocities are not defined. In \cite{Magpantay1}, it showed the velocities cannot be given in terms of the momenta, instead it yields a constraint which gives the Hamiltonian equal to zero. However for $ \alpha \neq 2 $, equation (7) yields the Hamiltonian
\begin{equation}\label{8}
\begin{split}
H_{r,\alpha}& = \sum_{a} \left[ \pi_{a} \dfrac{dx_{a 0}}{dt} + \vec{p}_{a} \cdot \dfrac{d\vec{x}_{a}}{dt} \right] - L_{r,\alpha}\\
                  & = \left[ \left(\frac{\alpha}{2}\right)^{\frac{\alpha}{2-\alpha}} - \left(\frac{\alpha}{2}\right)^{\frac{2}{2-\alpha}} \right] \left( \frac{-1}{mc} \right)^{\frac{\alpha}{2-\alpha}} \sum_{a} \left[ \pi_{a}^{2} - \vec{p}_{a} \cdot \vec{p}_{a} \right]^{\frac{1}{2-\alpha}}.
\end{split}
\end{equation}
The limit of $ H_{r,\alpha} $ as $ \alpha \rightarrow 2 $ is problematic. For example, just take the value $ \alpha = 1.99 $, the Hamiltonian becomes $ \propto \left[ \pi_{a}^{2} - \vec{p}_{a} \cdot \vec{p}_{a} \right]^{100}, $ a ridiculous momenta dependence for non-interacting relativistic particles. And as the accuracy of the $ \alpha $ limit is improved to 1.999 or worse 1.9999,  the Hamiltonian becomes $ \propto \left[ \pi_{a}^{2} - \vec{p}_{a} \cdot \vec{p}_{a} \right]^{1000} $ or worse $ \left[ \pi_{a}^{2} - \vec{p}_{a} \cdot \vec{p}_{a} \right]^{10000} $. Thus, breaking the reparametrization symmetry as given in equation (6) is not a viable way of getting a Hamiltonian in 8N phase space. 

If the relativistic two-body interaction term $ G_{4} $ is now included, the canonical procedure will now give the velocities in terms of the canonical momenta, now labeled by capital letters $ \Pi_{a}, \vec{P}_{a} $, as 
\begin{subequations}\label{9}
\begin{gather}
\dfrac{dx_{a0}}{dt} = \left( \frac{-1}{mc} \right)^{\frac{\alpha}{2-\alpha}} \left( \frac{\alpha}{2} \right)^{\frac{\alpha}{2-\alpha}} \left[ (\Pi_{a} + ...)^{2} - \vec{P}_{a} \cdot \vec{P}_{a} \right]^{\frac{\alpha-1}{2-\alpha}} (\Pi_{a} + ...),\label{first}\\
\dfrac{d\vec{x}_{a}}{dt} = - \left( \frac{-1}{mc} \right)^{\frac{\alpha}{2-\alpha}} \left( \frac{\alpha}{2} \right)^{\frac{\alpha}{2-\alpha}} \left[ (\Pi_{a} + ...)^{2} - \vec{P}_{a} \cdot \vec{P}_{a} \right]^{\frac{\alpha-1}{2-\alpha}} \vec{P}_{a}, 
\end{gather}
\end{subequations}
where the difference with equation (7) is that the $ G_{4} $ term changed $ \pi_{a} $ to $ \Pi_{a} + \frac{1}{2c^{2}} \sum_{b \neq a} \int dx_{b0} G_{4}(\vec{x}_{a} - \vec{x}_{b};\frac{1}{c}(x_{a0}-x_{b0})) $, and for compactness is simply written in equation (9) as $ (\Pi_{a} + ...) $. 
 
The Hamiltonian given in equation (8) becomes
\begin{equation}\label{10}
H_{r,\alpha} = \left[ \left(\frac{\alpha}{2}\right)^{\frac{\alpha}{2-\alpha}} - \left(\frac{\alpha}{2}\right)^{\frac{2}{2-\alpha}} \right] \left( \frac{-1}{mc} \right)^{\frac{\alpha}{2-\alpha}} \sum_{a} \left[ (\Pi_{a} + ...)^{2} - \vec{P}_{a} \cdot \vec{P}_{a} \right]^{\frac{1}{2-\alpha}}. 
\end{equation}
It also has the same problem as equation (8) when $ \alpha \rightarrow 2 $.  

5. It is for this reason that there is a need for breaking the reparametrization symmetry differently. This time consider
\begin{subequations}\label{11}
\begin{gather}
S_{r,\beta} = \int dt L_{r,\beta}, \label{first}\\
\begin{split}
L_{r,\beta}& = \sum_{a} \Big\lbrace (-mc) \left( \eta_{\mu \nu} \dfrac{dx_{a \mu}}{dt}  \dfrac{dx_{a \nu}}{dt} \right)^{\frac{1}{2}} + \beta \left( \eta_{\mu \nu} \dfrac{dx_{a \mu}}{dt}  \dfrac{dx_{a \nu}}{dt} \right)\\
& \quad - \frac{1}{2c^{2}} \dfrac{dx_{a 0}}{dt} \sum_{b \neq a} \int dt' \dfrac{dx_{b 0}}{dt'} G_{4}( \vec{x}_{a} - \vec{x}_{b}; \frac{1}{c} ( x_{a 0}(t) - x_{b 0}(t') ) )\Big\rbrace.
\end{split}
\end{gather}
\end{subequations} 
The $ beta = 0 $ gives the reparametrization symmetric action, which led to a constraint making the canonical Hamiltonian zero and thus no special relativistic Liouville equation. The $ \beta $ term, which breaks the reparametrization symmetry, is expected to not lead to a constraint and a non-zero Hamiltonian. The analogue of equation (7) for $ L_{r,\beta} $ without the two-body interaction $ G_{4} $ is
\begin{subequations}\label{12}
\begin{gather}
\dfrac{dx_{a0}}{dt} = \frac{1}{2\beta} \left[ 1 + mc\left( \pi_{a}^{2} - \vec{p}_{a} \cdot \vec{p}_{a} \right)^{\frac{-1}{2}} \right] \pi_{a},\label{first}\\
\dfrac{d\vec{x}_{a}}{dt} = - \frac{1}{2\beta} \left[ 1 + mc\left( \pi_{a}^{2} - \vec{p}_{a} \cdot \vec{p}_{a} \right)^{\frac{-1}{2}} \right] \vec{p}_{a}.
\end{gather}
\end{subequations}
The analogue of the free particle Hamiltonian in this case is
\begin{equation}\label{13}
\begin{split}
H_{r,\beta}& = \sum_{a} \left[ \pi_{a} \dfrac{dx_{a0}}{dt} + \vec{p}_{a} \cdot \dfrac{d\vec{x}_{a}}{dt} \right] - L_{r,\beta}\\
                & = \frac{1}{2\beta} \sum_{a} \left\lbrace \frac{3}{2} m^{2} c^{2} + 3 mc \left( \pi_{a}^{2} - \vec{p}_{a} \cdot \vec{p}_{a} \right)^{\frac{1}{2}} + \frac{3}{2} \left( \pi_{a}^{2} - \vec{p}_{a} \cdot \vec{p}_{a} \right) \right\rbrace.
\end{split} 
\end{equation}
Compared to equation (8), the dependence on the reparametrization breaking parameter $ \beta $, though singular in the limit 0, is rather simple. The Hamiltonian's phase space variable dependence is not affected by $ \beta $, thus the parameter $ \beta $ can simply just be scaled out. This is shown in the next section.

If the relativistic two-body interaction term $ G_{4} $ is included, the analogue of equation (9) is now
\begin{subequations}\label{14}
\begin{gather}
\dfrac{dx_{a0}}{dt} = \frac{1}{2\beta} \left\lbrace 1 + mc \left[ \left( \Pi_{a} + ... \right) ^{2} - \vec{P}_{a} \cdot \vec{P}_{a} \right]^{\frac{-1}{2}} \right\rbrace  \left( \Pi_{a} +...\right) \label{first}\\
\dfrac{d\vec{x}_{a}}{dt} = - \frac{1}{2\beta} \left\lbrace 1 + mc \left[ \left( \Pi_{a} + ... \right) ^{2} - \vec{P}_{a} \cdot \vec{P}_{a} \right]^{\frac{-1}{2}} \right\rbrace \vec{P}_{a},
\end{gather}
\end{subequations} 
where the term $ (\Pi_{a} + ...) $ above is the same as the term explained before equation (10).
The analogue of the Hamiltonian given in equation (10) is 
\begin{equation}\label{15}
\begin{split}
H_{r,\beta}&= \sum_{a} \left[ \Pi_{a} \dfrac{dx_{a0}}{dt} + \vec{P}_{a} \cdot \dfrac{d\vec{x}_{a}}{dt} \right] - L_{r,\beta}\\
                 & = \frac{1}{2\beta} \sum_{a} \left\lbrace \frac{3}{2} m^{2} c^{2} + 3 mc \left[  \left( \Pi_{a} + ...\right)^{2} - \vec{P}_{a} \cdot \vec{P}_{a} \right]^{\frac{1}{2}} + \frac{3}{2} \left[ \left( \Pi_{a} + ...\right)^{2} - \vec{P}_{a} \cdot \vec{P}_{a} \right] \right\rbrace. 
\end{split}                  
\end{equation} 
Just like the Hamiltonian given by equation (13), the beta parameter of the reparametrization symmetry breaking simply factors out as $ \frac{1}{\beta} $ and not in a complicated way as the $ \alpha $ parameter in equations (8) and (10). For this reason, it can be taken that the Hamiltonian in 8N phase space is given by equation (15) without the $ \frac{1}{\beta} $  factor as will be shown in the next section.

Since the special relativistic system in 8N phase space with one universal, Lorentz invariant time t has a canonical Hamiltonian, the Liouville equation exists and has the same form as the classical non-relativistic Liouville equation.

6. Now, a comparison is made between the Liouville equations given by the two ways the reparametrization symmetry is broken. In 8N phase space, the Liouville equation given by
\begin{equation}\label{16} 
\dfrac{\partial f}{\partial t} - \left\lbrace  H_{r}, f \right\rbrace  = 0,
\end{equation}
where the Gibbs distribution $ f = f(x_{0 a},\vec{x}_{a};\Pi_{a},\vec{P}_{a};t) $ with $ a = 1,...N $ and t is the Lorentz invariant universal time introduced in equation (3). Also, $ H_{r} $ is the relativistic Hamiltonian, which only exists if the reparametrization symmetry of t is explicitly broken. In the above discussions, this is done in two ways yielding the Hamiltonians given by equations (10) and equation (15). 

The Hamiltonian $ H_{r,\alpha} $ defined in equation (10) can be written as
\begin{subequations}\label{17} 
\begin{gather}
H_{r,\alpha} = \frac{1}{\gamma} \tilde{H}_{r,\alpha}, \label{first}\\
\gamma = \left[ \left(\frac{\alpha}{2}\right)^{\frac{\alpha}{2-\alpha}} - \left(\frac{\alpha}{2}\right)^{\frac{2}{2-\alpha}} \right]^{-1}, \label{second}\\
\tilde{H}_{r,\alpha} = (\frac{-1}{mc})^{\frac{\alpha}{2-\alpha}} \sum_{a} \left[ (\Pi_{a} + ...)^{2} - \vec{P}_{a} \cdot \vec{P}_{a} \right]^{\frac{1}{2-\alpha}},
\end{gather}
\end{subequations} 
where the Hamiltonian $ \tilde{H}_{r,\alpha} $ contains the phase space variables and physical constants and $ \gamma $ is a factor that goes with the choice of how the reparametrization symmetry is broken. The Liouville equation can now be written as
\begin{equation}\label{18}
\dfrac{\partial f}{\partial (\frac{t}{\gamma})} - \left\lbrace  \tilde{H}_{r,\alpha}, f \right\rbrace  = 0,
\end{equation}
Equation (18) gives the evolution of the Gibbs distribution for any value of $ \alpha
$. Note that since the Hamiltonian equations of motion can also be written similarly with $ t $ replaced by $ \frac{t}{\gamma} $ and $ H_{r,\alpha} $ replaced by $ \tilde{H}_{r,\alpha} $, then the Gibbs distribution function f has the following dependence $ f = f( x_{0 a},\vec{x}_{a};\Pi_{a},\vec{P}_{a};\frac{t}{\gamma} ) $. 

As $ \alpha $ is varied to approach the critical value equal to 2, the phase space variable dependence of $ \tilde{H}_{r,\alpha} \rightarrow \left[ (\Pi_{a} + ...)^{2} - \vec{P}_{a} \cdot \vec{P}_{a} \right]^{n} $ with n increasing fast and approaching $ \infty $ at $ \alpha = 2 $ while $ \gamma \rightarrow \infty $, which means $ \frac{t}{\gamma} \rightarrow 0 $. This means the Gibbs distribution when the reparametrization symmetry is restored is given by the initial condition defined by a very singular $ \tilde{H}_{r,\alpha} $. This does not make sense because the initial condition depends on how the system is prepared and not on the Hamiltonian that determines its dynamics.  

However, if the breaking of the reparametrization symmetry follows the approach given in Section 5, the phase space behaviour of $ H_{r,\beta} $ (see equation (15)) is the same regardless of the value of the breaking parameter $ \beta $. The analogue of equation (17) in this case is 
\begin{subequations}\label{19}
\begin{gather}
H_{r,\beta} = \frac{1}{\beta} \tilde{H}_{r}, \label{first}\\
\tilde{H}_{r} = \frac{1}{2} \sum_{a} \left\lbrace \frac{3}{2} m^{2} c^{2} + 3 mc \left[  \left( \Pi_{a} + ...\right)^{2} - \vec{P}_{a} \cdot \vec{P}_{a} \right]^{\frac{1}{2}} + \frac{3}{2} \left[  \left( \Pi_{a} + ...\right)^{2} - \vec{P}_{a} \cdot \vec{P}_{a} \right] \right\rbrace.
\end{gather}
\end{subequations}
The Liouville equation can now be written as 
\begin{equation}\label{20}
\dfrac{\partial f}{\partial (\frac{t}{\beta})} - \left\lbrace  \tilde{H}_{r}, f \right\rbrace  = 0,
\end{equation}
Since the Hamiltonian equations of motion can also be written similarly with $ t $ replaced by $ \frac{t}{\beta} $ and $ H_{r,\beta} $ replaced by $ \tilde{H}_{r} $, then the Gibbs distribution function f has the following dependence $ f = f( x_{0 a},\vec{x}_{a};\Pi_{a},\vec{P}_{a};\frac{t}{\beta} )$. The important difference here is that $ \tilde{H}_{r} $ is well-behaved and does not depend on $ \beta $. The $ \beta $ dependence is thus captured in the time-dependence being $ \frac{t}{\beta} $. Equation (20) then captures the non-equilibrium statistical mechanics of the N relativistic particles. 

As the parameter $ \beta $ varies and even as  the critical value $ \beta = 0 $ is approached, the Liouville equation is well-defined. Furthermore, $ \frac{t}{\beta} \rightarrow \infty $ at the critical value of $ \beta = 0 $, which means the Gibbs distribution when the reparametrization symmetry is fully restored is given by the equilibrium distribution defined by a well-behaved $ \tilde{H}_{r} $. It is for this reason that Liouville equation defined by breaking the reparametrization symmetry as given in Section 5 is a viable Special Relativistic Liouville Equation in 8N phase space.

\begin{acknowledgments}
I dedicate this paper to my apo, David, who gave me such happiness just seeing him smile, play and babble. 
\end{acknowledgments}

\end{document}